\documentclass{ws-ijmpd}
\usepackage[super,compress]{cite}
\usepackage{epstopdf}
\begin{document}

\markboth{M. F. A. R. Sakti, A. Suroso, and F. P. Zen}
{CFT Duals on Extremal Rotating NUT Black Holes}

%
\catchline{}{}{}{}{}
%

\title{CFT Duals on Extremal Rotating NUT Black Holes}

\author{Muhammad F. A. R. Sakti\footnote{corresponding author}}

\address{Theoretical Physics Lab., THEPI Division, Institut Teknologi Bandung, Jl. Ganesha 10\\
Bandung, 40132, Indonesia\\
$ ^* $m.fitrah@students.itb.ac.id}

\author{A. Suroso$ ^\dagger $ and Freddy P. Zen$ ^\ddagger $}

\address{Theoretical Physics Lab., THEPI Division, and Indonesia Center for Theoretical and Mathematical Physics (ICTMP), Institut Teknologi Bandung, Jl. Ganesha 10\\
Bandung, 40132, Indonesia\\
$ ^\dagger $agussuroso@fi.itb.ac.id\\
$ ^\ddagger $fpzen@fi.itb.ac.id}

\maketitle

\begin{history}
\received{Day Month Year}
\revised{Day Month Year}
\end{history}

\begin{abstract}
We investigate the Kerr-Newman-NUT black hole solution obtained from Pleba{\'n}ski-Demia{\'n}ski solutions with several assumptions. The origin of the microscopic entropy of this black hole is studied using the conjectured Kerr/CFT correspondence which is first proposed for extremal Kerr black holes. The isometry of the near-horizon extremal Kerr-Newman-NUT black hole shows that the asymptotic symmetry group may be applied to compute the central charge of the Virasoro algebra. Furthermore, by assuming Frolov-Thorne vacuum, the temperatures can be obtained which then using Cardy formula, the microscopic entropy is obtained and agrees with the Bekenstein-Hawking entropy. We also assume the case when the lowest eigenvalue of the conformal operator $ L_0 $ is non-zero to find the logarithmic correction of the entropy of NHEKNUT black hole. At the limit $ J \rightarrow 0 $, the extremal Reissner-Nordstr{\"o}m-NUT solution is produced and by adding the fibered coordinate we find the 5D solution. The second dual CFT is used to find the entropy and it still produces the area law of 5D black hole solution. So, the extremal Reissner-Nordstr{\"o}m-NUT solution is also holographically dual to the CFT.
\end{abstract}


\ccode{PACS numbers: 04.40.Nr, 04.60.-m, 04.70.Dy}


\section{Introduction}
\label{sec:intro}
Based on the no-hair theorem, the Kerr-Newman solution is the most physical black hole solution with the spin $ a $, mass $ M $, and the electric charge $e $. However, theo-retically there are many mathematical solutions that describe black holes, as one of the most common solutions, because they have various parameters other than those found in Kerr-Newman black hole, is Pleba{\'n}ski-Demia{\'n}ski (PD) solution \cite{plebanski1976}. The PD solution has additional parameters such as magnetic charge $ g $,  acceleration $ \alpha $, cosmological constant $ \Lambda $, and NUT charge $ l $. The presence of cosmological constants and acceleration in the solution adds the number of the horizon to the black hole known as accelerated and cosmological horizons, respectively. A simpler solution that we will discuss in the following is the case when the cosmological constant and acceleration vanish. So the solution becomes dyonic Kerr-Newman-NUT black hole. We will mention this solution as KNUT black hole.

The KNUT solution here is a more general solution compared to the dyonic Kerr-Newman one because there is an additional NUT charge which was first discovered by Taub \cite{taub1951} and later developed by Newman \textit{et al.} \cite{nut1963}. The interesting part of this solution is the presence of NUT charge in which the NUT is an abbreviation  of Newman, Unti, and Tamburinno that find the solution for larger manifold after Taub. The existence of a NUT charge causes the metric does not posses asymptotically flat solution. For the definition of the NUT charge itself, one can see it as the source that represents a gravomagnetic monopole parameter of the central mass \cite{nouri1997} or a twist parameter of the surrounding space-time \cite{badawi2006}.

Studying the thermodynamics quantities of KNUT black holes seems to be an interesting and important thing because it can be one of the stages to study the merging of gravitational theory and the quantum field theory. The conjectured Kerr/CFT correspondence \cite{guica2009} has first demonstrated that the microscopic entropy matches the (macroscopic) Bekenstein-Hawking entropy for extremal Kerr black hole. This shows that there is a relation between quantum gravity and the conformal field theory (CFT). In other words, from this conjecture, the gravity is dual to the CFT. This correspondence has also been successfully used to obtain the entropy of various black hole solutions from microscopic point of view such in Refs.  \citen{hartman2009,ghezelbash2009,lu2009,li2010,anninos2010, ghodsi2010,ghezelbash2012,astorino2015,astorino2015a, siahaan2016,astorino2016,sinamuli2016} or for the detailed review see Ref. \citen{compere2017}. We know that the entropy from the CFT is equal to $ A_H/4G_N $ where $ A_H, G_N $ are area of the black hole and gravitational constant in any dimension. This entropy clearly matches the Bekenstein-Hawking entropy.

Furthermore the macroscopic calculation in loop quantum gravity \cite{kaul2000} find that there is a logarithmic correction of the entropy , i.e. $ -\frac{3}{2}\text{ln}A_H $. Then, from the microscopic point of view, Cardy prescription succeed to show that logarithmic correction is consistent with the result obtained in the macroscopic calculation in loop quantum gravity \cite{carlip2000,pradhan2016}. This prescription assumes that the lowest eigenvalue of the conformal operator $ L_0 $ is non-zero. However, from the macroscopic side, Sen \cite{sen2012,sen2013} employs the analysis of Euclidean gravity to find the logarithmic correction and find that it depends on the ensemble that is chosen and the number of all massless fields in the theory. Thereafter, it is confirmed by Pathak \textit{et al.} \cite{pathak2017} where the logarithmic correction is generated by Gaussian fluctuations of the density of states about the saddle point from the microscopic side for Kerr-Newman black hole in five dimension. Sen argues that the loop quantum gravity result counts the number of states per unit interval in the area $ A_H $ but when converting this to number of states per unit interval in mass gives a result of $ -\text{ln} A_H $ for the singlet and no correction for the microcanonical ensemble in the case of Schwarzschild black hole. But he shows for the BTZ black hole, the logarithmic correction agrees with the microscopic one as the Cardy prescription. This logarithmic calculation seems to be very interesting for both assumption on macroscopic side and will be very promising to confirm it from the microscopic side.

In what follows, we use the Kerr/CFT correspondence for more generalized Kerr-Newman black holes, namely KNUT solution where this can be obtained from Pleba{\'n}ski-Demia{\'n}ski solution. First we get the near-horizon geometry of the KNUT black hole in some coordinate systems for which all of the metric have the $ SL(2,R) \times U(1) $ isometry group. From this symmetry, it might be seen that the asymptotic symmetry group (ASG), as proposed first in Ref. \citen{brown1986}, may be applied to Near-Horizon-Extremal-Kerr-Newman-NUT (NHEKNUT) geometry to obtain the central charge. Therefore, there are some boundary conditions on the metric deviation that need to be determined in order to produce a finite charge and here we focus only on the left-moving part of the central charge so that there are the additional boundary conditions to the charges. With the boundary condition we use, in the end this generates the central charge
\begin{eqnarray}
c_L = 12 aM ,\
\end{eqnarray}
which arise from Virasoro algebra. As we have said previously, here we assume that there is no contribution from the electromagnetic field to the central charge as well as the vanishing right-moving central charge.

After getting the associated central charge, the next step is to obtain the temperatures because these two quantities are required in Cardy formula for the entropy. In the assumptions we use, the temperature required in the computation of the entropy is the left-moving temperature. By using the Frolov-Thorne vacuum, we find that
\begin{eqnarray}
T_L = \frac{M^2 +(a+l)^2}{4\pi aM}.
\end{eqnarray}
But actually besides that, we also obtain the temperatures which is the conjugate of the electric and magnetic charges from KNUT black hole. It will be useful for finding the entropy for Reissner-Nordtr{\"o}m-NUT black hole in five dimensions. Finally, from Cardy formula, we find that the entropy of NHEKNUT black hole is
\begin{eqnarray}
S_{KNUT} = \pi (2a^2+q^2+2al), \
\end{eqnarray}
which agrees with the Bekenstein-Hawking entropy.

When the angular momentum of the NHEKNUT geometry vanishes, it causes the central charge to vanishes too but the left-moving temperature becomes singular at this point but this should still produce the entropy corresponding to the Bekenstein-Hawking entropy. To solve this problem, we adopt the method used by Hartmann \textit{et al.} \cite{hartman2009} which defines the second dual CFT. This is done by first adding a fibered coordinate that has a periodic property so that $ z \sim z + 2 \pi R_n $. In addition, the electromagnetic field obtained after adding the gauge transformation, because $ J \rightarrow 0 $, could be the part of the geometry so that a near-horizon metric of five dimensional extremal Reissner-Nordstrom-NUT solution can be obtained. Next, by using ASG and adding boundary conditions for the five dimensional solution, it produce the central charge of 5D geometry. Then by using the temperature that is the conjugate of the electric charge, the following entropy is obtained
\begin{eqnarray}
S_{RNNUT} =\pi (M^2+l^2)=\pi q^2. \
\end{eqnarray}
This entropy still agrees with the Bekenstein-Hawking entropy in five dimensional geometry where $ S_{BH} =A_H/4G_5 $.

\section{Kerr-Newman-NUT black holes}
\label{sec:kerr}
The no-hair theorem tells that the Kerr-Newman solution is the most physical black hole solution that has mass, electric charge, and spin. However, from the theoretical point of view, there are many mathematical solutions that describe black holes, as one of the general solutions, because they have various parameters other than those found in Kerr-Newman black hole, is Pleba{\'n}ski-Demia{\'n}ski (PD) solution \cite{plebanski1976}. This solution has the form \cite{podolsky2009}
\begin{eqnarray}
ds^2 &=& \frac{1}{\Omega ^2} \left[ -\frac{\Delta}{\Sigma} \left\{ d\hat{t} - \left(a \text{sin}^2\theta +4 l \text{sin}^2 \frac{\theta}{2}\right)d\hat{\phi} \right\}^2 +\frac{\Sigma}{\Delta}d\hat{r}^2 + \frac{\Sigma}{P}d\theta ^2 \right. \nonumber\\
&+& \left. \frac{P}{\Sigma}\text{sin}^2\theta \left\{ a d\hat{t} +\left(\hat{r}^2+(a+l)^2\right) d\hat{\phi} \right\}^2 \right],
\end{eqnarray}
where
\begin{eqnarray}
&& \Omega = 1- \frac{\alpha}{\omega}(l+a \text{cos}\theta)\hat{r}, ~ P = 1- P_0 \text{cos}\theta - P_1  \text{cos}^2\theta, ~ \Sigma = \hat{r}^2 + (l+a \text{cos}\theta)^2, \nonumber\\
&& \Delta= (\omega ^2 P_4 +e ^2 +g^2) - 2 M \hat{r}+P_2 \hat{r}^2 - 2\alpha \frac{P_3}{\omega}\hat{r}^3- \left( \alpha ^2 P_4 + \frac{\Lambda}{3} \right)\hat{r}^4, \nonumber\
\end{eqnarray}
and
\begin{eqnarray}
&& P_0 =2\alpha \frac{a}{\omega} M - 4\alpha ^2 \frac{a l}{\omega ^2} (\omega^2 P_4 +e ^2 +g^2) - 4\frac{\Lambda}{3}a l , \nonumber\\
&& P_1 = - \alpha ^2 \frac{a^2}{\omega ^2} (\omega^2 P_4 +e ^2 +g^2) - \frac{\Lambda}{3}a^2 ,\nonumber\\
&&  P_2 = \frac{\omega ^2 P_4}{a^2-l^2} +4\alpha \frac{l}{\omega} M - (a^2+3l^2)\left\{ \frac{\alpha^2}{\omega ^2}(\omega^2 P_4 +e ^2 +g^2) + \frac{\Lambda}{3} \right\} ,  \nonumber\\
&& P_3 = \frac{\omega ^2 l P_4}{a^2-l^2} -\alpha \frac{a^2 - l^2}{\omega} M + (a^2-l^2)l\left\{ \frac{\alpha^2}{\omega ^2}(\omega^2 P_4 +e ^2 +g^2) + \frac{\Lambda}{3} \right\} , \nonumber\\
&& P_4 = \left\{ 1+ 2\alpha \frac{l}{\omega}M-3\alpha ^2 \frac{l^2}{\omega ^2}(e ^2 +g^2)-l^2 \Lambda \right\} \left\{ \frac{\omega ^2}{a^2-l^2}+3\alpha ^2 l^2 \right\}^{-1} . \nonumber\
\end{eqnarray}
A detailed derivation of the solution can be seen in the original paper, Ref. \citen{plebanski1976}. The parameters $ M, a, e, g, l, \alpha, \Lambda $ represent mass, spin, electric charge, magnetic charge, NUT charge, acceleration, and cosmological constant, respectively where we use the notation $ k_B=\hbar=c=1 $ and the four dimensional gravitational constant also $ G=1 $. Then $ \omega $ is a free parameter that we can choose as a function of the previous seven parameters. In addition, the NUT charge can be defined as a gravomagnetic monopole parameter \cite{nouri1997} or a twisting property of the surrounding space-time \cite{badawi2006}. Its name comes from the abbreviation of Newman, Unti, and Tamburinno that find the solution for larger manifold \cite{nut1963} after Taub \cite{taub1951}.

In what follows, we focus on some special cases since we want only look for the Kerr-Newman-NUT solution. To find the KNUT solution, we take the value of the cosmologial constant $ \Lambda $ and acceleration $ \alpha $ to 0, so it results in $ P_0=P_1=0 $. Then the circumstances $ P_2 =1, P_3 =l, $ and $P_4=1 $ are taken. Finally, we obtain
\begin{eqnarray}
ds^2 &=& -\frac{\Delta}{\Sigma} \left\{ d\hat{t} - \left(a \text{sin}^2\theta +4 l \text{sin}^2 \frac{\theta}{2}\right)d\hat{\phi} \right\}^2 +\frac{\Sigma}{\Delta}d\hat{r}^2 + \Sigma d\theta ^2 \nonumber \\
&+& \frac{\text{sin}^2\theta}{\Sigma} \left[ a d\hat{t} +\left\{\hat{r}^2+(a+l)^2\right\} d\hat{\phi} \right]^2, \ \label{eq:KerrNewmanNUT0}
\end{eqnarray}
where
\begin{eqnarray}
\Sigma = \hat{r}^2 + (l+a \text{cos}\theta)^2,~ \Delta=\hat{r}^2 - 2 M \hat{r}+  q^2 +a^2 -l^2. \nonumber\
\end{eqnarray}
Here, we have defined $ q^2 =e^2 +g ^2 $. The solution (\ref{eq:KerrNewmanNUT0}) is equivalent to the following solution
\begin{eqnarray}
ds^2 &=& -\frac{\Delta}{\Sigma} \left[ d\hat{t} - \left\{a \text{sin}^2\theta +2 l (1-\text{cos}\theta) \right\}d\hat{\phi} \right]^2 +\frac{\Sigma}{\Delta}d\hat{r}^2 + \Sigma d\theta ^2  \nonumber \\
&+& \frac{\text{sin}^2\theta}{\Sigma} \left\{ a d\hat{t} +\left(\hat{r}^2+a^2+l^2+2al \right) d\hat{\phi} \right\}^2. \label{eq:KerrNewmanNUT}
\end{eqnarray}
The metric (\ref{eq:KerrNewmanNUT}) is related to the electromagnetic potential \cite{podolsky2009}
\begin{eqnarray}
A_\mu dx^\mu &=& \frac{-e\hat{r}\left[a d\hat{t}- \left\{ (a+l)^2 - (l^2+a^2 \text{cos}^2\theta +2al \text{cos}\theta) \right\} d\hat{\phi} \right]}{a\left[\hat{r}^2 + (l+a\text{cos}\theta)^2 \right]}  \nonumber\\
&-& \frac{g(l+a\text{cos}\theta)\left[a d\hat{t}- \left\{ \hat{r}^2 + (l+a\text{cos}\theta)^2 \right\} d\hat{\phi} \right]}{a\left[\hat{r}^2 + (l+a\text{cos}\theta)^2 \right]} .\ \label{eq:electromagneticpotential}
\end{eqnarray}

The metric (\ref{eq:KerrNewmanNUT}) has the event horizon, angular velocity, and Hawking temperature, respectively
\begin{eqnarray}
\hat{r}_+ &=& M + \sqrt{M^2 +l^2 -a^2 - q^2}, \\
\Omega_H &=& \frac{a}{ \left[ \hat{r}_+^2+(a+l)^2 \right] } ,\\
T_H &=& \frac{\kappa}{2\pi} = \frac{r_+ - M}{2\pi \left[ \hat{r}_+^2+(a+l)^2 \right]} ,\label{eq:TH} \
\end{eqnarray}
in which the angular velocity is the same as with the accelerating black hole Kerr-Newman-NUT  \cite{jan2014}. When its acceleration parameter vanishes, the Hawking temperature will be equal to Eq. (\ref{eq:TH}). For the extremal case, it deals with Bekenstein-Hawking entropy
\begin{eqnarray}
S_{BH} (T_H=0) = \frac{A_H}{4} = \pi (2a^2+q^2+2al). \label{eq:entropiBH}
\end{eqnarray}

However, there is a logarithmic correction to the Bekenstein-Hawking entropy in the in loop quantum gravity theory \cite{kaul2000}. In this theory, the macroscopic calculation finds that the logarithmic correction of the entropy of the black hole has the form
\begin{eqnarray}
\Delta S_{BH} = -\frac{3}{2} \text{ln} A_H . \label{eq:entropyBHcorrection} \
\end{eqnarray}
Hence for the near-horizon extremal KNUT black hole, the logarithmic correction of the entropy is
\begin{eqnarray}
\Delta S_{BH} = -\frac{3}{2} \text{ln}  4\pi (2a^2+q^2+2al) .\
\end{eqnarray}
Beside the result from the loop quantum gravity, the logarithmic correction can also be obtained from the analysis of Euclidean gravity such in Refs. \citen{sen2012} and \citen{sen2013}. Therein the logarithmic correction is found that depends on the ensemble that is chosen and the number of all massless fields in the theory. 

\section{Near-horizon geometry of Kerr-Newman-NUT black holes}
\label{sec:Near-horizon}

In this section, we wish to study the near-horizon geometry of extremal Kerr-Newman-NUT space-time where the extremal case occurs when $ M^2=a^2+q^2-l^2 $. In order to do so, we change the coordinates by the following transformations \cite{hartman2009,compere2017}
\begin{eqnarray}
\hat{r} = \hat{r}_+ +\lambda r_0 y, ~~~\hat{t} = \frac{r_0}{\lambda}\tau , ~~~ \hat{\phi} = \varphi + \Omega_H \frac{r_0}{\lambda}\tau ,\label{eq:translocal}\
\end{eqnarray}
where we choose $ r_0=(2a^2+q^2+2al)/M $ and the parameter $ \lambda $ approaches zero. The metric (\ref{eq:KerrNewmanNUT}) changes then to the near-horizon metric given by
\begin{eqnarray}
ds^2= \chi(\theta) \left(-\frac{r_0^2}{M^2} y^2 d\tau ^2 +\frac{dy^2}{y^2}+d\theta ^2 \right) + \frac{r_0 ^2 M^2 \text{sin}^2 \theta}{\chi(\theta)} \left( d\varphi + \frac{2a}{M}y d\tau \right)^2 , \label{eq:NHEKNUT}
\end{eqnarray}
where $ \chi(\theta) = r_+^2 +(l+a \text{cos}\theta)^2 = a^2(1+\text{cos}^2\theta) +2al\text{cos}\theta +q^2 $. This is the Near-Horizon Extremal Kerr-Newman-NUT (NHEKNUT) metric. In the near-horizon limit $ \lambda \rightarrow 0 $ or degenerate horizon, the electromagnetic potential is not regular. In this case, such problem can be circumvented by expanding the electromagnetic potential \cite{bicak2015} but first we need to introduce the Coulomb electromagnetic potential
\begin{eqnarray}
\Phi_H = - K^\mu A_\mu \big|_{\hat{r}=\hat{r}_+} = \frac{e \hat{r}_+}{\hat{r}_+^2+(a+l)^2}, \label{eq:coulombpotential}
\end{eqnarray}
where $ K = \partial _{\hat{t}}+\Omega_H \partial_{\hat{\phi}} $. The Coulomb electromagnetic potential will be used as the gauge transformation in the electromagnetic potential. By expanding the electromagnetic potential (\ref{eq:electromagneticpotential}) in $ r-r_+ = \lambda r_0 y $ and add the gauge transformation (see \ref{sec:appendixa}), we obtain
\begin{eqnarray}
A_\mu dx^\mu = f(\theta) \left(d\varphi +\frac{2a}{M}yd\tau \right)-\frac{e[M^2-(a+l)^2]}{2aM}d\varphi ,\ \label{eq:nearhorizonelectromagneticpot}
\end{eqnarray}
where
\begin{eqnarray}
f(\theta) = \frac{\left[M^2+(a+l)^2 \right]\left[2gMl+e(M^2-l^2)+a(2gM-2el-ae\text{cos}\theta)\text{cos}\theta \right]}{2aM \left[M^2+(a\text{cos}\theta+l)^2 \right]}. \label{eq:ftheta} \
\end{eqnarray}
Note that this metric (\ref{eq:NHEKNUT}) and the electromagnetic potential (\ref{eq:nearhorizonelectromagneticpot}) will remain the same as the near-horizon geometry of Kerr-Newman space-time when the NUT charge vanishes.

We can see clearly that the metric (\ref{eq:NHEKNUT}) has time-like Killing vector, so we can remove the constant over $ y^2 d\tau^2 $ by using this scaling
\begin{eqnarray}
yd\tau \rightarrow \frac{M}{r_0} yd\tau .\label{eq:scalingtimelike}
\end{eqnarray}
Because of the scaling (\ref{eq:scalingtimelike}), the metric (\ref{eq:NHEKNUT}) is then
\begin{eqnarray}
ds^2= \chi(\theta) \left(- y^2 d\tau ^2 +\frac{dy^2}{y^2}+d\theta ^2 \right) + \frac{r_0 ^2 M^2 \text{sin}^2 \theta}{\chi(\theta)} \left( d\varphi + ky d\tau \right)^2 , \label{eq:NHEKNUT1}
\end{eqnarray}
where $ k= 2aM/(2a^2+q^2+2al) $. Because of the scaling (\ref{eq:scalingtimelike}), the near-horizon electromagnetic potential (\ref{eq:nearhorizonelectromagneticpot}) becomes
\begin{eqnarray}
A_\mu dx^\mu = f(\theta) \left(d\varphi +k yd\tau \right)-\frac{c}{k}d\varphi ,\ \label{eq:nearhorizonelectromagneticpotential1}
\end{eqnarray}
where $ c=e[M^2 -(a+l)^2]/[M^2 +(a+l)^2]$ and this is the general form of near-horizon electromagnetic field as the one discussed in Ref. \citen{compere2017}, but the second term in the right hand side can be gauged away as in Ref.  \citen{hartman2009}.

Besides the metric form (\ref{eq:NHEKNUT1}), the NHEKNUT geometry can be represented in Poincar{\'e}-type coordinates such used in some papers of the Kerr/CFT correspondence \cite{guica2009,ghezelbash2009}. The Poincar{\'e}-type of metric of the NHEKNUT geometry can be obtained by using the following transformations
\begin{eqnarray}
\hat{r} = \hat{r}_+ +\frac{r_0\lambda}{\hat{y}}, ~~~\hat{t} &=& \frac{r_0}{\lambda}\hat{\tau}, ~~~ \hat{\phi} = \hat{\varphi} + \Omega_H \frac{r_0}{\lambda}\hat{\tau} ,\label{eq:TransfPoincare}\
\end{eqnarray}
and followed by the scaling
\begin{eqnarray}
\frac{d\hat{\tau}}{\hat{y}} \rightarrow \frac{M}{r_0} \frac{d\hat{\tau}}{\hat{y}}, \label{eq:scalingtimelike1}
\end{eqnarray}
to the metric (\ref{eq:KerrNewmanNUT}). The Poincar{\'e}-type of NHEKNUT geometry is then
\begin{eqnarray}
ds^2= \chi(\theta) \left(\frac{-d\hat{\tau} ^2 + d \hat{y}^2}{\hat{y}^2}+d\theta ^2 \right) + \frac{r_0 ^2 M^2 \text{sin}^2 \theta}{\chi(\theta)} \left( d\hat{\varphi} + k \frac{d\hat{\tau}}{\hat{y}} \right)^2 . \label{eq:NHEKNUTPoincare}
\end{eqnarray}
The near-horizon electromagnetic potential now has the form
\begin{eqnarray}
A_\mu dx^\mu = f(\theta) \left(d\varphi +\frac{kd\tau}{y} \right)-\frac{c}{k}d\varphi .\ \label{eq:nearhorizonelectromagneticpotential2}
\end{eqnarray}

The metrics (\ref{eq:NHEKNUT}), (\ref{eq:NHEKNUT1}), and (\ref{eq:NHEKNUTPoincare}) are not asymptotically flat but asymptotically Anti-de Sitter (AdS). Those metrics cover only part of the NHEKNUT geometry. To cover the whole near-horizon geometry, we use the global coordinates. In order to find the global form, we just need to transform the metrics (\ref{eq:NHEKNUT1}) or (\ref{eq:NHEKNUTPoincare}). The metric (\ref{eq:NHEKNUTPoincare}), by the following global coordinate transformation \cite{guica2009,ghezelbash2012,bardeen1999,sakti2016,sakti2018a}
\begin{eqnarray}
\hat{y}=\frac{1}{r+\sqrt{1+r^2}\mathrm{cos}t}, ~ \hat{\tau}=\hat{y} ~ \mathrm{sin}t \sqrt{1+r^2},~ \hat{\varphi} =\phi + k \mathrm{ln}\left(\frac{\mathrm{cos}t +r~\mathrm{sin}t}{1+\mathrm{sin}t \sqrt{1+r^2}}\right), \label{eq:nheknutglobalransf}
\end{eqnarray}
will transform the Poincar{\'e}-type of NHEKNUT geometry, such that we obtain
\begin{eqnarray}
ds^2= \chi(\theta) \left[-(1+r^2)dt ^2 + \frac{dr^2}{1+r^2}+d\theta ^2 \right] + \frac{r_0 ^2 M^2 \text{sin}^2 \theta}{\chi(\theta)} \left( d\phi + k rdt \right)^2 . \label{eq:NHEKNUTglobal}
\end{eqnarray}
This global form can also be obtained from the metric (\ref{eq:NHEKNUT1}) by the similar transformations such in Ref.  \citen{mei2010}. The global form of near-horizon electromagnetic field is then
\begin{eqnarray}
A_\mu dx^\mu = f(\theta) \left(d\phi +k rdt \right). \label{eq:globalgauge} \
\end{eqnarray}
Note that we have added the gauge transformation before applying the global coordinate transformations to the near-horizon electromagnetic field.

For a fixed polar angle $ \theta $, the near-horizon geometry is a quotient of warped AdS$ _3 $ which the quotient arises from identification of $ \phi $ coordinate. This near-horizon geometry of KNUT black hole has $SL(2,R)\times U(1) $ isometry group where for the global NHEKNUT one, the subgroup $ U(1) $ is generated by the Killing vector 
\begin{eqnarray}
\zeta_0 = -\partial_\phi, \
\end{eqnarray}
and the subgroup $SL(2,R)$ is generated by the three Killing vectors
\begin{eqnarray}
&&X_0=2\partial_t, \nonumber\\
&&X_1 = 2\mathrm{sin}t \frac{r}{\sqrt{1+r^2}}\partial_t-2\mathrm{cos}t \sqrt{1+r^2} \partial_r+\frac{2 \mathrm{sin}t}{\sqrt{1+r^2}}\partial_\phi ,\nonumber\\
&&X_2 = -2\mathrm{cos}t \frac{r}{\sqrt{1+r^2}}\partial_t-2\mathrm{sin}t \sqrt{1+r^2}\partial_r-\frac{2 \mathrm{cos}t}{\sqrt{1+r^2}}\partial_\phi .\
\end{eqnarray}
This is a hint that tells us, according to the conjectured Kerr/CFT correspondence, the near-horizon extremal black holes could be dual to the CFT and the asymptotic symmetry group (ASG) may be applied. 

\section{Asymptotic Symmetry Group}
\label{sec:ASG}
We now employ the approach of Brown and Henneaux \cite{brown1986} to find the central charge of the holographic dual conformal field theory description of an extremal rotating black hole. Because the KNUT black hole is a solution of Einstein-Maxwell theory, it seems that there exists the non-vanishing contributions to the central charge from electromagnetic field besides the metric tensor. First, we assume the non-zero angular momentum $ J $. Later in the section \ref{sec:second} we will see the case when $ J=0 $ and the second dual CFT such used in Ref. \citen{hartman2009} will be used to compute the entropy.

\subsection{Charges}
To compute the charges associated with asymptotic symmetry group (ASG) of near horizon extremal Kerr-Newman-NUT black hole, we should consider all possible contributions from all different fields in the action and we can use the formalism in Ref. \citen{barnich2002}. Asymptotic symmetries of this solution include diffeomorphisms $ \xi $ such that
\begin{eqnarray}
&& \delta_\xi A_\mu = \mathcal{L}_\xi A_\mu = \xi^\mu (\partial_\nu A_\mu) + A_\nu (\partial_\mu \xi^\nu) , \\
&& \delta_\xi g_{\mu\nu} = \mathcal{L}_\xi g_{\mu\nu} =  \xi^\sigma (\partial_\sigma g_{\mu\nu}) + g_{\mu\sigma}(\partial _\nu \xi^\sigma)+ g_{\sigma \nu}(\partial _\mu \xi^\sigma), \ 
\end{eqnarray}
as well as the following $ U(1) $ gauge transformation $ \Lambda $
\begin{eqnarray}
\delta_\Lambda A_\mu = \partial_\mu \Lambda .
\end{eqnarray}
We denote the metric deviation and the electromagnetic field as  $ \delta_\xi A_\mu = a_\mu $ and $ \delta_\xi g_{\mu\nu} = h_{\mu\nu} $. So there are two contributions to the associated charge of the asymptotic symmetry group from the Kerr-Newman-NUT solution, which is the contribution of the metric tensor and the electromagnetic field. The associated conserved charge is given by
\begin{eqnarray}
Q_{\xi,\Lambda} = \frac{1}{8\pi}\int_{\partial\Sigma} \left(k^{g}_{\zeta}[h;g]+k^{A}_{\zeta,\Lambda}[h,a;g,A]  \right),
\end{eqnarray}
where the integral is over the boundary of a spatial slice. The explicit expressions for the contribution of the metric tensor and electromagnetic field on the central charge respectively are\begin{eqnarray}
k^{g}_{\zeta}[h;g] &=& -\frac{1}{4}\epsilon_{\rho\sigma\mu\nu} \left\{ \zeta ^{\nu} D^{\mu} h - \zeta ^{\nu} D_{\lambda} h^{\mu \lambda} + \frac{h}{2} D^{\nu}\zeta^{\mu} - h^{\nu \lambda} D_{\lambda}\zeta^{\mu} + \zeta_{\lambda}D^{\nu}h^{\mu \lambda} \right. \nonumber\\
& & \left. +\frac{h^{\lambda \nu}}{2}\left(D^\mu \zeta_\lambda + D_\lambda \zeta^\mu \right) \right\} dx^\rho \wedge dx^\sigma , \label{eq:kgrav} \\ 
k^{A}_{\zeta,\Lambda}[h,a;g,A] &=& \left[ \epsilon_{\alpha\beta \mu\nu} \left\{\frac{1}{8} \left( -\frac{1}{2}hF^{\mu\nu}+2F^{\mu\gamma}h_\gamma^\nu - \delta F^{\mu\nu} \right) (\zeta^\rho A_\rho + \Lambda) -\frac{1}{8}F^{\mu\nu}\zeta^\rho a_\rho \right. \right. \nonumber\\ 
& &\left. \left. -\frac{1}{4} F^{\alpha \mu}\zeta^\nu a_\alpha  \right\} - \frac{1}{8}\epsilon_{\alpha \beta}^{\mu \nu}a_\mu (\mathcal{L}_\zeta A_\nu + \partial_\nu \Lambda) \right] dx^\alpha \wedge dx^\beta , \label{eq:kgauge}\
\end{eqnarray}
where $ \delta F^{\mu\nu} \equiv g^{\mu \alpha} g^{\nu \beta}(\partial_\alpha a_\beta - \partial_\beta a_\alpha) $. We should note that the last two terms in Eq. (\ref{eq:kgrav}) as well as in Eq. (\ref{eq:kgauge}) vanish for an exact Killing vector and an exact symmetry, respectively.

The charge $ Q_{\zeta,\Lambda} $ generates symmetry through the Dirac brackets. The algebra of the asymptotic symmetric group is given by the Dirac bracket algebra of these charges
\begin{eqnarray}
\{ Q_{\zeta,\Lambda},Q_{\bar{\zeta},\bar{\Lambda}} \}_{DB} &=& \frac{1}{8\pi}\int \left(k^{g}_{\zeta}\left[\mathcal{L}_{\bar{\zeta}}g;g \right]+k^{A}_{\zeta,\Lambda}\left[\mathcal{L}_{\bar{\zeta}}g,\mathcal{L}_{\bar{\zeta}}A+d\bar{\Lambda};g,A \right]  \right) \nonumber\\
&=& Q_{[(\zeta,\Lambda),(\bar{\zeta},\bar{\Lambda})]} + \frac{1}{8\pi}\int \left(k^{g}_{\zeta}\left[\mathcal{L}_{\bar{\zeta}}\bar{g};\bar{g}\right]+k^{A}_{\zeta,\Lambda}\left[\mathcal{L}_{\bar{\zeta}}\bar{g},\mathcal{L}_{\bar{\zeta}}\bar{A}+d\bar{\Lambda};\bar{g},\bar{A}\right]  \right), \label{eq:charges}\nonumber\\
\
\end{eqnarray}

\subsection{Boundary Conditions}
To use ASG, we need to specify the boundary conditions of the metric and the electromagnetic potential deviations. The boundary conditions are imposed to produce finite charges for both gravitational and electromagnetic parts. Therefore, we adopt the boundary conditions such in most Kerr/CFT correspondence articles for both fields. For the metric deviation of the global metric form, we impose the following boundary conditions
\begin{eqnarray}
h_{\mu \nu} \sim \left(\begin{array}{cccc}
\mathcal{O}(r^2) & \mathcal{O}\left(\frac{1}{r^2}\right) &  \mathcal{O}\left(\frac{1}{r}\right) &  \mathcal{O}(1) \\
 &  \mathcal{O}\left(\frac{1}{r^3}\right) &  \mathcal{O}\left(\frac{1}{r^2}\right) &  \mathcal{O}\left(\frac{1}{r}\right) \\
 &  & \mathcal{O}\left(\frac{1}{r}\right) &  \mathcal{O}\left(\frac{1}{r}\right)\\
 &  &  &  \mathcal{O}(1)\
\end{array} \right), \label{eq:gdeviation}
\end{eqnarray}
in the basis $ (t,r,\theta, \phi) $. Then for the electromagnetic potential, we impose the following boundary conditions
\begin{eqnarray}
a_\mu \sim \left( \mathcal{O}\left( r\right), \mathcal{O}\left( 1/r^2 \right),  \mathcal{O}\left( 1\right) , \mathcal{O}\left( 1/r\right) \right).
\end{eqnarray}
As in Ref. \citen{hartman2009}, we append the additional boundary condition because we want to focus only on the central charge that comes from the left-moving Virasoro algebra. So we impose the boundary condition
\begin{eqnarray}
Q_{\partial_t} = 0 .\
\end{eqnarray}
In this case, we want the total central charge due to the diffeomorphism transformation of the fields comes from the gravity only, as in the case of near-horizon extremal Kerr-Newman-AdS black holes. So another boundary condition is imposed, i.e.
\begin{eqnarray}
Q_{\Lambda} = 0 .\
\end{eqnarray}
Hence as we have denoted, in the Dirac bracket calculation of the conserved charges (\ref{eq:charges}), we consider the central term which is a result of the left-moving part only.

\subsection{Central charge}
The most general diffeomorphism symmetry that preserves such boundary conditions ({\ref{eq:gdeviation}) is generated by the vector field
\begin{eqnarray}
\zeta &=& \left\{c_t + \mathcal{O}\left(r^{-3}\right) \right\}\partial _t + \left\{-r\epsilon '(\phi) + \mathcal{O}(1) \right\}\partial _r + \mathcal{O}\left(r^{-1}\right)\partial _\theta \nonumber\\ & & + \left\{\epsilon(\phi) + \mathcal{O}\left(r^{-2}\right) \right\}\partial _\phi ,\
\end{eqnarray}
where $ c_t $ is an arbitrary constant and the prime $ (') $ denotes the derivative respect to $ \phi $. This asymptotic symmetry group contains one copy of the conformal group of the circle generated by
\begin{eqnarray}
\zeta_\epsilon = \epsilon(\phi)\partial_\phi - r\epsilon '(\phi)\partial_r ,\label{eq:killingASG}
\end{eqnarray}
that will be the part of the NHEKNUT metric. We know that the azimuthal coordinate is periodic under the rotation $ \phi \sim \phi+2\pi $ hence we may define $ \epsilon_n = -e^{-in \phi} $ and $ \zeta_\epsilon =\zeta_\epsilon(\epsilon_n ) $. Because of that, the generator of ASG (\ref{eq:killingASG}) becomes
\begin{eqnarray}
\zeta_\epsilon = -e^{-in \phi}\partial _\phi - inre^{-in \phi}\partial_r . \label{eq:killingASG1}
\end{eqnarray}
By the Lie bracket, the symmetry generator (\ref{eq:killingASG1}) satisfy the Virasoro algebra
\begin{eqnarray}
i[\zeta_m, \zeta_n]_{LB} = (m-n)\zeta_{m+n},
\end{eqnarray}
without the central term because we do not define the quantum version yet. The non-zero metric deviations are
\begin{eqnarray}
&& h_{tt} = 2inr^2 \left[\frac{\chi ^2(\theta)- 4a^2M^2 \text{sin}^2\theta}{\chi(\theta)} \right]e^{-in\phi} ,\nonumber\\
&& h_{rr} = \frac{-2in\chi(\theta)}{(1+r^2)^2}e^{-in\phi}, ~~~ h_{r\phi} = \frac{-n^2 r \chi(\theta)}{1+r^2}e^{-in\phi}, \\
&& h_{\phi \phi} = \frac{2in(2a^2 +q^2 +2al)^2 \text{sin}^2\theta}{\chi(\theta)} e^{-in\phi} .\nonumber\
\end{eqnarray}
The Dirac brackets of the conserved charges are now just the common forms of the Virasoro algebras with a central term
\begin{eqnarray}
\{ Q_{\zeta},Q_{\bar{\zeta}} \}_{DB} = Q_{[\zeta,\bar{\zeta}]} + \frac{1}{8\pi}\int k^{g}_{\zeta}\left[\mathcal{L}_{\bar{\zeta}}\bar{g};\bar{g}\right] . \label{eq:chargesbracket}\
\end{eqnarray}
Then by defining
\begin{eqnarray}
Q_{\zeta} \equiv L_{n} - \varrho \delta_{n,0},
\end{eqnarray}
where $ \varrho = 3aM/2 $ in this case, we obtain the conserved charges algebra in quantum version, such that
\begin{eqnarray}
\left[L_m, L_n \right] = (m-n) L_{m+n} + aM m (m^2-1)\delta_{m+n, 0}. 
\end{eqnarray}
From the algebra above, we can read-off the value of the left-moving central charge for NHEKNUT black hole, i.e.
\begin{eqnarray}
c_L = 12aM = 12 a\sqrt{a^2 +q^2 -l^2}. \label{eq:cLNHEKNUT}\
\end{eqnarray}

\section{Temperature}
\label{sec:temperature}
After getting the central charge in the previous section, we need to find the temperature in order to use Cardy formula to obtain the entropy. To find the temperature, we use the analog of the Hartle-Hawking vacuum, i.e. Frolov-Thorne vacuum that have been used in the Kerr/CFT correspondence because the angular momentum is included within this vacuum. When the Hawking temperature is zero, this vacuum is a pure state. But here will be a little bit difference comparing to \cite{guica2009} because there is an emergence of the additional thermodynamics potentials such as electric and magnetic potentials. Both potentials are the conjugate of the electric and magnetic charges, respectively. Now, we may use the first law of the black hole thermodynamics
\begin{eqnarray}
T_H dS = dM - \Omega_H dJ - \Phi_e dQ_e - \Phi_g dQ_g ,
\end{eqnarray}
and the extremal condition that satisfies
\begin{eqnarray}
T_H^{ex} dS = dM - \Omega_H^{ex} dJ - \Phi_e^{ex} dQ_e - \Phi_g^{ex} dQ_g =0 .
\end{eqnarray}
Therefore we can obtain
\begin{eqnarray}
T_H dS = -\left[ (\Omega_H - \Omega_H^{ex})dJ + (\Phi_e - \Phi_e^{ex}) dQ_e + (\Phi_g - \Phi_g^{ex})dQ_g \right]. \label{eq:constrainofthermo}
\end{eqnarray}
The electric potential and magnetic potential can be defined from (\ref{eq:electromagneticpotential}), which are given by
\begin{eqnarray}
\Phi_e &=& \frac{e\hat{r}_++g(a+l)}{\hat{r}_+^2+(a+l)^2}, \label{eq:electricpot} \\
\Phi_g &=&  \frac{-g(a+l)}{\hat{r}_+^2+(a+l)^2},  \label{eq:magneticpot}\
\end{eqnarray}
where the extremal case means $ \hat{r}_+ = M $. The potentials (\ref{eq:electricpot}) and (\ref{eq:magneticpot}) will be useful in the derivation of the temperatures related to the electromagnetic charges.
For such constrained variations (\ref{eq:constrainofthermo}), we may write
\begin{eqnarray}
dS = \frac{dJ}{T_L}+\frac{dQ_e}{T_e}+\frac{dQ_g}{T_g} .\
\end{eqnarray}

We consider the quantum scalar field with eigenmodes of the asymptotic energy $ E $ and angular momentum $ J $, which are given by the following form
\begin{eqnarray}
\tilde{\Phi} = \sum_{E,J,s} \tilde{\phi} _{E,J,s} e^{-i E \hat{t} + i J \hat{\phi}} f_s(\hat{r},\theta),
\end{eqnarray}
for Kerr black hole. In order to transform this to near-horizon quantities and take the extremal limit, we note that in the near-horizon coordinates (\ref{eq:translocal}) we have
\begin{eqnarray}
e^{-i E \hat{t} + i J \hat{\phi}} = e^{-i (E-\Omega_H^{ex} J )\tau r_0/\lambda + i J \varphi} = e^{-in_R \tau + in_L \varphi},
\end{eqnarray}
where
\begin{eqnarray}
n_R = -(E-\Omega_H^{ex} J)r_0/\lambda, ~~~n_L = J. \label{eq:nrnl}
\end{eqnarray}
But this is only suitable when there is no contribution of the electromagnetic potential. So in our case, using the fact that there are potentials as the conjugates of the electric and magnetic potentials, we may extend Eq. (\ref{eq:nrnl}) to 
\begin{eqnarray}
n_R = -(E-\Omega_H^{ex} J- \Phi_e^{ex} Q_e- \Phi_g^{ex} Q_g) r_0/\lambda, ~~~n_L = J. \label{eq:nRnLnew}
\end{eqnarray}
Hence the density matrix in the asymptotic energy, angular momentum, electric charge, and magnetic charge eigenbasis now has the Boltzmann weighting factor
\begin{eqnarray}
e^{-\left( \frac{E - \Omega_H J - \Phi_e^{ex} Q_e- \Phi_g^{ex} Q_g}{T_H}\right) } =e^{-\frac{n_R}{T_R}-\frac{n_L}{T_L}-\frac{Q_e}{T_e}-\frac{Q_g}{T_g} }. \label{eq:Boltzmannweightingfactor}\
\end{eqnarray}

If we take the trace over the modes inside the horizon, the Boltzmann weighting factor will be a diagonal matrix. We can compare the Eqs. (\ref{eq:nRnLnew}) and (\ref{eq:Boltzmannweightingfactor}) to obtain the definition of the temperatures of the CFT, such that
\begin{eqnarray}
&&T_R = \frac{T_H r_0}{\lambda}\bigg|_{ex} , ~~~ T_L = - \frac{\partial T_H/\partial \hat{r}_+}{\partial \Omega_H / \partial \hat{r}_+}\bigg|_{ex}, \nonumber\\
&&T_e = - \frac{\partial T_H/\partial \hat{r}_+}{\partial \Phi_e / \partial \hat{r}_+}\bigg|_{ex}, ~~~ T_g = - \frac{\partial T_H/\partial \hat{r}_+}{\partial \Phi_g / \partial \hat{r}_+}\bigg|_{ex} .\ \label{eq:generalCFTtemperature}
\end{eqnarray}
We know that for the extremal black holes, the Hawking temperature $ T_H $ will vanish but not all of the temperatures from the CFT will also vanish. We can see that only the right-moving temperature which is equal to zero in the extremal case and the others finally become
\begin{eqnarray}
&& T_L = \frac{M^2 +(a+l)^2}{4\pi aM}, \nonumber\\
&& T_e = \frac{M^2 +(a+l)^2}{2\pi \left[ 2gM(a+l)+e \left\{M^2 -(a+l)^2 \right\} \right]}, \label{eq:temperatureofNHEKNUT}\\
&& T_g = -\frac{M^2 +(a+l)^2}{4\pi gM(a+l)}. \nonumber\
\end{eqnarray}
We can also write the left-moving temperature in the form $ T_L = 1/2\pi k $ where $ k $ is constant that is obtained in section \ref{sec:Near-horizon}.  The Hartle-Hawking vacuum state is generalized around the extremal KNUT black hole with a density matrix given by
\begin{eqnarray}
\rho = e^{-\frac{J}{T_L}-\frac{Q_e}{T_e}-\frac{Q_g}{T_g}}. \label{eq:mixedstate}\
\end{eqnarray}
Because of the boundary CFT is dual to the NHEKNUT black hole, the dual of this black hole is described by the CFT in the mixed state (\ref{eq:mixedstate}).

\section{Entropy from CFT}
\label{sec:entropy}
\subsection{Cardy formula}
To compute the entropy, we know use the famous Cardy formula that comes from the CFT. This entropy has the general form
\begin{eqnarray}
S = 2\pi \left( \sqrt{\frac{c_L E_L}{6}} + \sqrt{\frac{c_R E_R}{6}} \right),\label{eq:cardyentropi}\
\end{eqnarray}
where $ E_L,E_R $ are the eigen energies of the operator $ L_0, \bar{L}_0 $. In order to obtain the entropy as a function of temperature as used in the Kerr/CFT correspondence, we use the thermodynamic relation $ dE = T dS $, such that
\begin{eqnarray}
S = \frac{\pi ^2}{3}(c_L T_L+c_R T_R),\label{eq:cardyentropi1}\
\end{eqnarray}
For the NHEKNUT solution, there are also $ T_e $ and $ T_g $ that actually have to be taken into account to the Cardy entropy. But in section \ref{sec:ASG}, we have assumed that the charge from the electromagnetic field contribution is zero. Because the right-moving central charge is also zero and by using the central charge (\ref{eq:cLNHEKNUT}) and left-moving temperature in Eq. (\ref{eq:temperatureofNHEKNUT}), the microscopic entropy of NHEKNUT black hole is then
\begin{eqnarray}
S = \frac{\pi ^2}{3}c_L T_L = \pi (2a^2+q^2+2al) =\frac{A_H}{4} .\label{eq:cardyentropi2}\
\end{eqnarray}
It precisely agrees with the macroscopic Bekenstein-Hawking entropy (\ref{eq:entropiBH}). This entropy reduces to entropy of dyonic Kerr-Newman for $ l=0 $, Kerr-NUT for $ q=0 $, and Kerr for $ l,q=0 $, so this entropy is more general.

\subsection{Logarithmic correction}

Here we will show the logarithmic corrrection of the entropy as shown in Ref. \citen{kaul2000} especially for the NHEKNUT black hole. The general form of this logarithmic correction from the CFT is shown in Refs. \citen{carlip2000} and \citen{pradhan2016} from Cardy prescription that we use here. In Cardy formula, the corrections come from the addition of the lowest eigenvalue $ E_ {L0} $ of the conformal operator $ L_0 $, that often but not always has a zero value. The choice of boundary conditions actually affects the result, but in general as it is imposed in the section \ref{sec:ASG}, the resulting Virasoro algebra generates the central charge and the eigenvalue of $ L_0 $ such as the following, respectively
\begin{eqnarray}
c_L = \frac{3 A_H}{2\pi G_N}\frac{\beta}{\kappa}, ~ E_L = \frac{A_H}{16\pi G_N} \frac{\kappa}{\beta} ,
\end{eqnarray}
where $ \kappa $ is a surface gravity and $ \beta $ is an undetermined periodicity. It is clearly seen that we still only focus on the left-moving part and we have assumed $ E_{L0} << c_L $ in the derivation. Thus the logarithmic correction of the entropy that depends on the area of the black hole is given by
\begin{eqnarray}
\Delta S = - \frac{3}{2}\text{ln} A_H. \label{eq:entropycorrection}\
\end{eqnarray}

If we take into account this logarithmic correction to NHEKNUT metric, we will have
\begin{eqnarray}
\Delta S = -\frac{3}{2} \text{ln}   4\pi (2a^2+q^2+2al) . \label{eq:correctedentropyNHEKNUT} \
\end{eqnarray}

For $ q=0 $, Eq. (\ref{eq:correctedentropyNHEKNUT}) will be the correction of extremal Kerr-NUT black hole \cite{sakti2018b}. Note that the microscopic logarithmic correction obtained here come from the assumption of the non-zero lowest eigenvalue of the conformal operator and based on the macroscopic result in loop quantum gravity \cite{kaul2000}. We emphasize this because in Ref. \citen{pathak2017}, they find the logarithmic correction that is generated by Gaussian fluctuations of the density of states about the saddle point to confirm the macroscopic result from \cite{sen2013} that employ the analysis of Euclidean gravity. Actually to go further, we may calculate the logarithmic correction such in Ref. \citen{pathak2017}. However we need to find the macroscopic result before to confirm the microscopic one because in Refs. \citen{sen2012} and \citen{sen2013}, it is not assumed that there is NUT charge in the black hole solution that actually come from the gravity.

\section{Reissner-Nordstr{\"o}m-NUT solution in 5D}
\label{sec:second}
In the NHEKNUT metric we find, when the momentum angular $ J $ approaches 0, it causes the central charge to vanish as well because its value is proportional to $ J $ and produces the extremal Reissner-Nordstr{\"o}m-NUT solution while the left-moving temperature will be singular because it is proportional to $ 1/J $. It seems like what happen to Kerr-Newman-AdS black hole. However, the resulting microscopic entropy also remains in accordance with the Bekenstein-Hawking entropy because the angular momentum in the central charge and temperature cancel each other. We do not want this to happen, so we need another way to get the microscopic entropy from the finite central charge and temperature.

As mentioned earlier, Hartman \textit{et al.} \cite{hartman2009} face the same case and manage to obtain the finite value of the microscopic entropy by using the second dual CFT. Here, we will adopt the same method as they do, which is actually also done in Ref. \citen{ghodsi2010}. First we need to add a new coordinate representing the fifth dimension of $ S ^1 $ which has property $ z \sim z + 2 \pi R_n $ where $ R_n $ is an integer. The addition of this fibered coordinate produces one additional Killing vector $ \partial_z $ which becomes $ U (1) $ symmetry in addition to $ SL (2, R)_R \times U (1)_L $. Furthermore, the electromagnetic potential can be mapped to a geometric part by a duality transformation such in string theory but it is not always applied \cite{guica2009a}. This is actually such a black hole solution that exists in Kaluza-Klein theory \cite{horowitz2011} where dimensional reduction from 5D to 4D will produce dilaton and electromagnetic fields in the four dimensional metric. The holographic dual of Kaluza-Klein black hole is also studied in Ref. \citen{azeyanagi2009}.

When the limit $ J \rightarrow 0 $, not only the left-moving temperature will be singular but also the electromagnetic potential but this singularity can be eliminated by taking a certain gauge transformation. To obtain the entropy of the five dimensional extremal Reissner-Nordstr{\"o}m-NUT black hole, we use the global form metric (\ref{eq:NHEKNUTglobal}) included its electromagnetic potential (\ref{eq:globalgauge}). For the electromagnetic potential in this solution, we choose the following gauge transformation
\begin{eqnarray}
\textbf{A} \rightarrow \textbf{A} - \frac{2gMl+e(M^2-l^2)}{2aM} d\phi .\
\end{eqnarray}
Hence when we take $ a=0 $, the electromagnetic potential is then given by
\begin{eqnarray}
\textbf{A}=A_\mu dx^\mu = k r dt + \frac{(gM-el)\text{cos}\theta}{M} d\phi . \label{eq:elecpotRNNUT} \
\end{eqnarray}
where for Reissner-Nordstr{\"o}m-NUT solution, the constant $ k = [2gMl+e(M^2-l^2)]/(M^2+l^2)$. This electromagnetic potential (\ref{eq:elecpotRNNUT}) can be the part of the geometry of the Reissner-Nordstr{\"o}m-NUT solution as we have said in preceding paragraph. Furthermore, the metric now is in the form
\begin{eqnarray}
ds_5^2 = ds_4^2 + (dz + \textbf{A})^2, \label{eq:RNNUT} \
\end{eqnarray}
where the four dimensional metric is
\begin{eqnarray}
ds_4^2= \chi \left[-(1+r^2)dt ^2 + \frac{dr^2}{1+r^2}+d\theta ^2 \right] + \frac{r_0 ^2 M^2 \text{sin}^2 \theta}{\chi} d\phi  ^2 , \label{eq:RNNUTglobal}
\end{eqnarray}
that we obtain from (\ref{eq:NHEKNUTglobal}) and taking $ a=0 $ and herein the section $ r_0 = q^2/M $ and $ \chi = M^2 +l^2 $. As mentioned in the preceding few paragraphs, this method is used to obtain the extremal Reissner-Nordstr{\"o}m-AdS geometry. Next, we will prove that the entropy of this black hole, by using the second dual CFT, will produce the following Bekenstein-Hawking entropy
\begin{eqnarray}
S_{BH} = \pi q^2 = \frac{A_H}{4G_5}.\
\end{eqnarray}

\subsection{Central charge}
Just like in the section \ref{sec:ASG}, we do need to compute the central charge to find the entropy as well as the temperature KNUT one where the central charge comes from the five dimensional gravity only, so we can neglect the contribution of the electromagnetic potential. We have to note that in 5D gravity, the gravitational constant is $ G_5 = 2 \pi $.

To find the non-trivial diffeomorphisms, we need to set some consistent boundary conditions of the metric deviation as in 4D solution. Here we adopt the same boundary condition such in Ref. \citen{hartman2009}, i.e.
\begin{eqnarray}
h_{\mu \nu} \sim \left(\begin{array}{ccccc}
\mathcal{O}(r^2) & \mathcal{O}\left(\frac{1}{r^2}\right) &  \mathcal{O}\left(\frac{1}{r}\right) & \mathcal{O}(r) &  \mathcal{O}(1) \\
 &  \mathcal{O}\left(\frac{1}{r^3}\right) &  \mathcal{O}\left(\frac{1}{r^2}\right) & \mathcal{O}\left(\frac{1}{r}\right) &  \mathcal{O}\left(\frac{1}{r}\right) \\
 &  & \mathcal{O}\left(\frac{1}{r}\right) & \mathcal{O}(1) &  \mathcal{O}\left(\frac{1}{r}\right)\\
 &  &  & \mathcal{O}\left(\frac{1}{r}\right) & \mathcal{O}(1)\\
 &  &  &  & \mathcal{O}(1)\
\end{array} \right), \label{eq:gdeviation2}
\end{eqnarray}
in the basis $ (t,r,\theta, \phi, z) $. Hence the most general diffeomorphisms are of the form
\begin{eqnarray}
\zeta &=& \left\{b_t + \mathcal{O}\left(r^{-3}\right) \right\}\partial _t + \left\{-r\epsilon '(z) + \mathcal{O}(1) \right\}\partial _r + \mathcal{O}\left(r^{-1}\right)\partial _\theta \nonumber\\ & & + \left\{b_{\phi} + \mathcal{O}\left(r^{-2}\right) \right\}\partial _\phi + \left\{ \epsilon(z) + \mathcal{O}\left(r^{-2}\right) \right\}\partial _z ,\
\end{eqnarray}
where $ \epsilon (z) = -e^{-inz} $ and the prime $ (') $ denotes the derivative respect to $ z $. Here $ b_t, b_\phi $ are arbitrary constants. We can see that the boundary conditions (\ref{eq:gdeviation2}) allow
\begin{eqnarray}
\zeta_z = \epsilon(z) \partial_z - r \epsilon '(z)\partial _z ,\
\end{eqnarray}
but do not allow $ \zeta_\epsilon $ such the case of 4D NHEKNUT geometry. Furthermore, to compute the central charge, we may follow the same steps such in section \ref{sec:ASG} but for five dimensional gravity \cite{ghodsi2010}, we have
\begin{eqnarray}
k^{g}_{\zeta}[h;g] &=& -\frac{1}{2}\frac{1}{3!}\epsilon_{\rho\sigma \gamma \mu\nu} \left\{ \zeta ^{\nu} D^{\mu} h - \zeta ^{\nu} D_{\lambda} h^{\mu \lambda} + \frac{h}{2} D^{\nu}\zeta^{\mu} - h^{\nu \lambda} D_{\lambda}\zeta^{\mu} + \zeta_{\lambda}D^{\nu}h^{\mu \lambda} \right. \nonumber\\
& & \left. +\frac{h^{\lambda \nu}}{2}\left(D^\mu \zeta_\lambda + D_\lambda \zeta^\mu \right) \right\} dx^\rho \wedge dx^\sigma \wedge dx^\gamma , \label{eq:kgrav2} \
\end{eqnarray}
where the last two terms vanish for the exact Killing vector. Finally, the central charge is given by
\begin{eqnarray}
c_z = 6k\chi, 
\end{eqnarray}
that is associated to $ \zeta_z $.

\subsection{Temperature}
We are left only by the temperatures conjugate to electric and magnetic charges. Hence from the first law of black hole thermodynamics, we have
\begin{eqnarray}
dS = \frac{dQ_e}{T_e}+\frac{dQ_g}{T_g}.
\end{eqnarray}
As the second dual CFT for extremal Reissner-Nordstr{\"o}m-AdS solution, the magnetic charge is held fixed. So now the vacuum is a pure state. The remaining temperature is given by
\begin{eqnarray}
T_e =\frac{1}{2\pi k} =\frac{M^2 +l^2}{2\pi \left[ 2gMl+e (M^2 -l^2) \right]}. \label{eq:temperatureofRNNUT}\
\end{eqnarray}
This can easily be obtained by taking $ a=0 $ in the temperature (\ref{eq:temperatureofNHEKNUT}).

\subsection{Entropy}
Finally, we can compute the microscopic entropy of the extremal Reissner-Nordstr{\"o}m-NUT black hole by using Cardy formula. The entropy of the extremal Reissner-Nordstr{\"o}m-NUT black hole is then
\begin{eqnarray}
S_{RNNUT} = \frac{\pi ^2}{3}c_z T_e =\pi (M^2+l^2)= \pi q^2 = \frac{A_H}{4G_5}.
\end{eqnarray}
Therefore the entropy from the CFT for this 5D solution also matches the Bekenstein-Hawking entropy. For zero NUT charge, it reduces to the entropy of extremal Reissner-Nordstr{\"o}m black hole \cite{ghodsi2010}. If we take into account the logarithmic correction from the Cardy prescription, the logarithmic correction for this black hole is
\begin{eqnarray}
\Delta S = -\frac{3}{2} \text{ln} 4\pi q^2 . \label{eq:correctedentropyRNNUT} \
\end{eqnarray}

In Ref. \citen{pathak2017}, there is a microscopic calculation of the logarithmic correction of 5D Kerr-Newman black hole where the correction is $ -4 \text{ln} d $, based on the macroscopic result from the analysis of Euclidean gravity \cite{sen2012,sen2013} where $ d $ is a length parameter that has relation $ d^{N-2}=A_H $. In the analysis of Euclidean gravity, there is a constant $ C_{local} $ in the correction that will vanish for five dimensional solution. For the 5D extremal Reissner-Nordstr{\"o}m-NUT black hole, that constant must vanish too and the rest of the constant depends on the ensemble that is chosen.

\section{Concluding Remarks}
\label{sec:concluding}
In this letter, we have investigated the duality between the NHEKNUT black hole and the CFT on its boundary by using the Kerr/CFT correspondence. We choose this black hole solution because it has NUT charge that makes the solution is not asymptotically flat and it is more general then the Kerr-Newman solution. In order to study the duality, first we have obtained the NHEKNUT solution in three types of coordinate which all of this solutions have the same isometry, i.e. $ SL(2,R)\times U(1) $, so ASG can be applied to find the central charge. Before computing the microscopic entropy, the Frolov-Thorne vacuum is considered and extended to add the additional contribution of the electromagnetic potential to obtain the temperatures. Finally, it is found that the entropy of NHEKNUT black hole is
\begin{eqnarray}
S = \pi (2a^2+q^2+2al),\nonumber\
\end{eqnarray}
that agrees with the Bekenstein-Hawking entropy and we have proved once again that the Kerr/CFT correspondence is true at this point. The point is also that this entropy is more general than the dyonic Kerr-Newman. Then we have showed also the logarithmic correction of the entropy coming from the CFT for this black hole by using the Cardy prescription.

An interesting case occurs when the angular momentum $ J $ vanishes. It makes NHEKNUT solution becomes the extremal Reissner-Nordstr{\"o}m-NUT solution and produces finite entropy but with the singular left-moving temperature and zero central charge. It makes us use the second dual CFT to prove that it still have to produce finite and the non-zero central charge and temperature to obtain the same entropy with the Bekenstein-Hawking one. We add the fifth fibered coordinate to extend the solution to 5D geometry and make the use of the electromagnetic potential as the part of the geometry too. After getting the 5D solution, we use the same method as we use for the NHEKNUT solution to find the central charge and the temperature then computing the entropy. At the end, we prove that the microscopic entropy is
\begin{eqnarray}
S_{RNNUT} = \pi (M^2+l^2)=\pi q^2 , \nonumber\
\end{eqnarray}
that comes from the second dual CFT, matches the Bekenstein-Hawking entropy in general and this entropy is more general than the Reissner-Nordstr{\"o}m one. The logarithmic correction from the Cardy prescription is also shown. 

The origin of NUT charge is still interesting to be understood and herein we know that Kerr/CFT correspondence do succeed again to study the microscopic origin of the entropy of the rotating NUT black holes. But in the other side, the microscopic calculation of the logarithmic correction of the entropy from the CFT is still debatable and needs to be confirmed with the macroscopic result especially for black holes that contain NUT charge.

As future work, we want to study the Kerr/CFT correspondence for the rotating black hole solution with quintessential dark energy. This solution adds a new parameter that represents the existence of quintessential dark energy around the black hole. We think that it maybe useful in astrophysics since there exist the dark energy.

\section*{Acknowledgments}

We gratefully acknowledge support from Ministry of Research, Technology, and Higher Education of the Republic of Indonesia and PMDSU Research Grant 2017. M.F.A.R.S. also thanks all members of Theoretical Physics Laboratory, Institut Teknologi Bandung for the valuable support.

\appendix
\section{Electromagnetic potential for the degenerate horizon}
\label{sec:appendixa}
For the degenerate horizon, the electromagnetic potential is not regular. To solve it, we can expand the electromagnetic potential such that
\begin{eqnarray}
\textbf{A} &=& A_\mu dx^\mu= A_{\hat{t}}d\hat{t} + A_{\hat{\phi}}d\hat{\phi} \nonumber\\
&=& \left(A_{\hat{t}}\big|_{\hat{r}=\hat{r}_+} + \frac{\partial A_{\hat{t}}}{\partial \hat{r}} \bigg|_{\hat{r}=\hat{r}_+} \lambda r_0 y \right) \frac{r_0}{\lambda}d\tau + \left(A_{\hat{\phi}}\big|_{\hat{r}=\hat{r}_+} + \frac{\partial A_{\hat{\phi}}}{\partial \hat{r}} \bigg|_{\hat{r}=\hat{r}_+} \lambda r_0 y \right) \times \nonumber\\
& &\left(d\varphi + \Omega_H \frac{r_0}{\lambda}d\tau \right) \nonumber\\
&=& -\Phi_H \frac{r_0}{\lambda}d\tau + \left(\frac{\partial A_{\hat{\phi}}}{\partial \hat{r}} \bigg|_{\hat{r}=\hat{r}_+} + \Omega_H \frac{\partial A_{\hat{t}}}{\partial \hat{r}} \bigg|_{\hat{r}=\hat{r}_+} \right)r_0 ^2 yd\tau  \nonumber \\
&  &+ \left(A_{\hat{\phi}}\big|_{\hat{r}=\hat{r}_+} + \frac{\partial A_{\hat{\phi}}}{\partial \hat{r}} \bigg|_{\hat{r}=\hat{r}_+} \lambda r_0 y \right) d\varphi .\
\end{eqnarray}
So if we want to take the limit $ \lambda \rightarrow 0 $, we have to use the gauge transformation
\begin{eqnarray}
\textbf{A} \rightarrow \textbf{A} + \Phi_H \frac{r_0}{\lambda}d\tau ,\
\end{eqnarray}
to remove the singularity.


\end{document}